\documentclass[sigconf]{acmart}
\usepackage{array}
\usepackage{pifont}
\usepackage{multirow}
\usepackage{multicol}
\usepackage{makecell}
\usepackage{subfig}
\usepackage{algorithm}
\usepackage{algorithmic}
\usepackage{booktabs}
\usepackage{soul}
\usepackage{color}
\usepackage{graphicx}
\usepackage{colortbl}
\usepackage{amsmath,amsfonts,amsthm,bm}

\newtheoremstyle{definitionstyle}%
  {3pt}
  {3pt}
  {\itshape} 
  {}
  {\bfseries}
  {:}
  {0.5em}
  {}
\theoremstyle{definitionstyle}

\newtheoremstyle{assumptionstyle}%
  {3pt}
  {3pt}
  {}
  {}
  {\bfseries}
  {:}
  {0.5em}
  {}

\theoremstyle{assumptionstyle}

\AtBeginDocument{%
  }

\copyrightyear{2026}
\acmYear{2026}
\setcopyright{cc}
\setcctype{by}
\acmConference[KDD 2026] {Proceedings of the 32nd ACM SIGKDD Conference on Knowledge Discovery and Data Mining V.2}{August 9--13, 2026}{Jeju Island, Republic of Korea.}
\acmBooktitle{Proceedings of the 32nd ACM SIGKDD Conference on Knowledge Discovery and Data Mining V.2 (KDD 2026), August 9--13, 2026, Jeju Island, Republic of Korea}
\acmISBN{979-8-4007-2259-2/2026/08}
\acmDOI{10.1145/3770855.3817625}
\acmSubmissionID{v2rtp0059}
\begin{document}

\title{LoRAShield: Data-Free Editing Alignment for Secure Personalized LoRA Sharing}

\author{Jiahao Chen}
\orcid{0000-0002-5894-662X}
\authornote{Equal contribution.}
\affiliation{%
  \institution{Zhejiang University}
  \department{College of Computer Science and Technology}
  \city{Hangzhou}
  \country{China}
}
\email{xaddwell@zju.edu.cn}

\author{Junhao Li}
\orcid{0009-0008-1233-2236}
\authornotemark[1]
\affiliation{%
  \institution{Zhejiang University}
  \department{College of Computer Science and Technology}
  \city{Hangzhou}
  \country{China}
}
\email{junhao.li@zju.edu.cn}

\author{Yiming Wang}
\orcid{0009-0005-3978-6797}
\affiliation{%
  \institution{Zhejiang University}
  \department{College of Computer Science and Technology}
  \city{Hangzhou}
  \country{China}
}
\email{ym_wang@zju.edu.cn}

\author{Yong Yang}
\orcid{0000-0003-3526-560X}
\affiliation{%
  \institution{Zhejiang University}
  \department{College of Computer Science and Technology}
  \city{Hangzhou}
  \country{China}
}
\email{yangyong2022@zju.edu.cn}

\author{Yi Jiang}
\orcid{0009-0007-2677-4062}
\affiliation{%
  \institution{Zhejiang University}
  \department{College of Computer Science and Technology}
  \city{Hangzhou}
  \country{China}
}
\affiliation{%
  \institution{Guizhou Medical University}
  \department{School of Biology and Engineering}
  \city{Guiyang}
  \country{China}
}
\email{jiangyi2021@zju.edu.cn}

\author{Chunyi Zhou}
\orcid{0000-0003-0081-0946}
\affiliation{%
  \institution{Zhejiang University}
  \department{College of Computer Science and Technology}
  \city{Hangzhou}
  \country{China}
}
\email{zhouchunyi@zju.edu.cn}

\author{Qingming Li}
\orcid{0000-0002-6085-7300}
\affiliation{%
  \institution{Zhejiang University}
  \department{College of Computer Science and Technology}
  \city{Hangzhou}
  \country{China}
}
\email{liqm@zju.edu.cn}

\author{Tianyu Du}
\orcid{0000-0003-0896-0690}
\affiliation{%
  \institution{Zhejiang University}
  \department{School of Software Technology}
  \city{Ningbo}
  \country{China}
}
\email{zjradty@zju.edu.cn}

\author{Shouling Ji}
\orcid{0000-0003-4268-372X}
\authornote{Corresponding author.}
\affiliation{%
  \institution{Zhejiang University}
  \department{College of Computer Science and Technology}
  \city{Hangzhou}
  \country{China}
}
\affiliation{%
  \institution{Zhejiang University}
  \department{Zhejiang Key Laboratory of Decision Intelligence}
  \city{Hangzhou}
  \country{China}
}
\email{sji@zju.edu.cn}

\renewcommand{\shortauthors}{Jiahao Chen et al.}

\ccsdesc[500]{Computing methodologies~Artificial intelligence}
\ccsdesc[300]{Security and privacy~Software and application security}

\keywords{diffusion model, text-to-image, adversarial attack, concept erasure}

\begin{abstract}
The proliferation of Low-Rank Adaptation (LoRA) has democratized personalized text-to-image generation, enabling users to share lightweight models (e.g., personal portraits) on platforms like Civitai and Liblib. However, this ``share-and-play'' ecosystem introduces critical but unnoticed risks: benign LoRAs can be weaponized by adversaries to generate harmful content (e.g., political, defamatory imagery), undermining creator rights and platform safety. 
To bridge this gap, we propose LoRAShield, the first data-free editing framework for securing LoRA models against misuse. Our platform-driven approach dynamically edits and realigns LoRA's weight subspace via adversarial optimization and semantic augmentation. Experimental results demonstrate that LoRAShield achieves remarkable effectiveness, efficiency, and robustness in blocking malicious generations without sacrificing the functionality of the benign task. By shifting the defense to platforms, LoRAShield enables secure, scalable sharing of personalized models, a critical step toward trustworthy generative ecosystems.
\textbf{Warnings}: \textit{This paper contains sexually and bloody explicit imagery that some readers may find disturbing, distressing, and/or offensive. To mitigate the offensiveness to readers, we showcase some of the explicit images with black masks.}
\end{abstract}


\maketitle

\section{Introduction}
\label{sec:intro}
The rapid development of text-to-image (T2I) technology~\cite{dhariwal2021diffusion,podell2023sdxl}, particularly diffusion models (DMs)~\cite{rombach2022high}, has dramatically improved personalized creativity. However, this advancement has simultaneously exacerbated problems like bias~\cite{luccioni2023stable,everaert2024exploiting,chen2026customizationfirepluginpoisoning}, copyright infringement~\cite{somepalli2023understanding}, and the generation of not-safe-for-work (NSFW) content~\cite{yang2024mma,yang2024sneakyprompt}. Malicious users of T2I services may exploit these models to create misleading images that harm individuals' interests or reputations or even provoke social panic~\cite{qu2023unsafe}. For instance, a finance worker at a multinational firm was tricked into paying out \$25 million to fraudsters using deepfake technology to pose as the company’s chief financial officer in a video conference call~\cite{deepfakecfo}.

\begin{figure}[t]
    \centering
    \subfloat[Misuse example 1]{\includegraphics[width=0.45\linewidth]{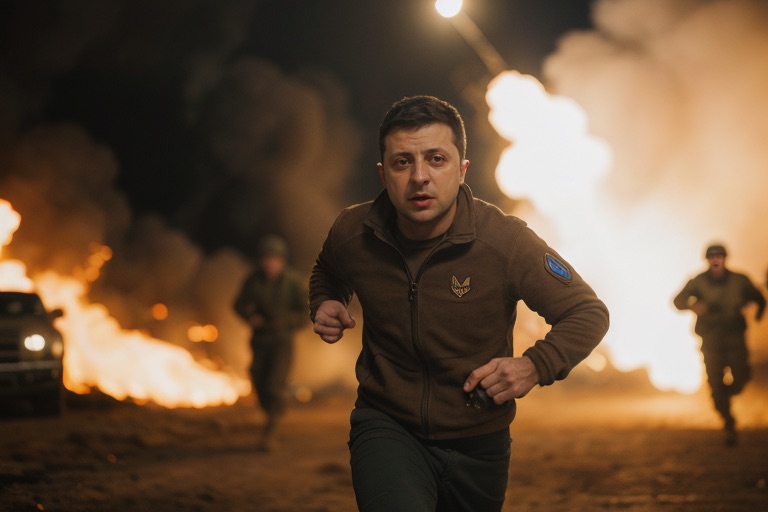}}
    \hspace{10pt}
    \subfloat[Misuse example 2]{\includegraphics[width=0.45\linewidth]{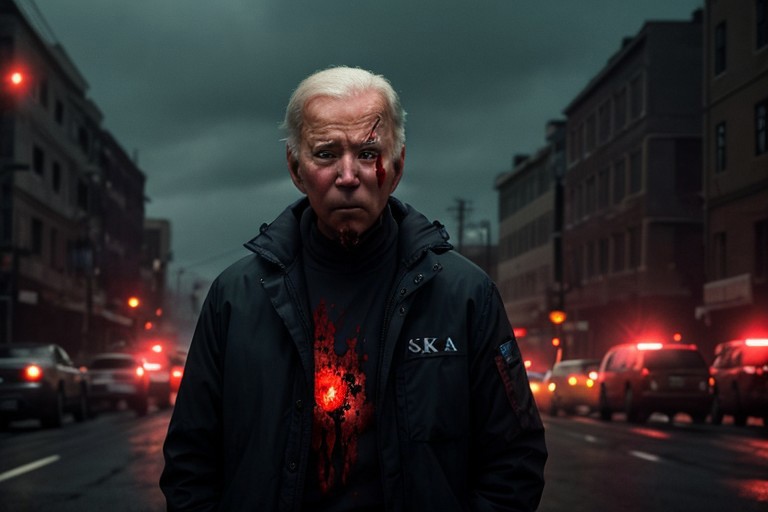}}
    \caption{Generated images using LoRA~\cite{joe-biden,zelenskyy} on Civitai. Note that these images will never be posted online.}
    \label{fig:misuse example}
\end{figure}

To mitigate such risks, numerous studies~\cite{abs-2412-18123,wu2024erasediff,gandikota2024unified,wang2025aeiouunifieddefenseframework} have explored the elimination of specific harmful concepts from DMs. However, existing research focuses mainly on the DMs~\cite{zhang2024unlearncanvas}, overlooking significant vulnerabilities in their plugins. In particular, Low-Rank Adaptation (LoRA)~\cite{hu2022lora} enables parameter-efficient adaptation of DMs, allowing users to customize models with minimal computational overhead. The widespread adoption of LoRA allows individuals to efficiently build personalized models from their local datasets (personal portraits or artworks). Users often share their LoRA models, instead of the entire DM, on platforms like Civitai~\cite{civitaimodels}, LiblibAI~\cite{liblib}, and Hugging Face~\cite{huggingface}. Any user can then download the shared LoRAs and incorporate them into their own image generation pipeline. Unfortunately, this share-and-play characteristic also introduces significant risks. Malicious users can use public LoRAs to generate harmful content. For example, our trial in Fig.~\ref{fig:misuse example} using Civitai’s online generation services reveal that despite the platform’s claims of preventing the generation of NSFW content~\cite{civitaiprotocol}. \textbf{Any benign LoRAs uploaded by genuine users could be exploited to produce content that can significantly disrupt their interests and reputation.} Thus, malicious actors can download benign LoRAs to generate harmful images, negatively impacting the original creators' interests or reputation. This situation indicates a critical gap in effective oversight by both academia and industry regarding the abuse of LoRAs.

To address this risk, we re-examined the threat models (Sec~\ref{sec:threat model}) associated with LoRA misuse in current model marketplaces and wondered if it is possible to develop a defense. Given the high volume of LoRA models uploaded to platforms~\cite{wei2024exploring}, the defense method must be lightweight and efficient, avoiding negative impacts on the quality of legitimate image generation. Additionally, we find that even though previous studies~\cite{zhang2024unlearncanvas,wu2024erasediff} have indeed proposed methods for concept erasure in DMs, they all failed to be efficient and data-free to adapt to practical scenarios. They exhibit instability, incomplete removal of target concepts, limited generalization, and robustness restricted to specific prompts (details in Sec.~\ref{sec:concept erasing}).

Further, we propose a platform-driven and data-free LoRA editing alignment method (LoRAShield) that allows users to specify undesired concepts, thus preventing their LoRAs from being exploited for malicious purposes. Specifically, users upload LoRAs to a platform that offers dynamic editing services based on user-defined constraints. By editing and realigning the LoRA's weight subspace, LoRAShield restricts the model's ability to produce certain undesired content while preserving the quality of its intended generation tasks for legitimate uses. In essence, our work aims to build the first critical line of defense before this vulnerability is exploited at scale. This philosophy is best captured by the adage: \textbf{``A stitch in time saves nine!''}. 

Our contributions are summarized as follows:
\begin{itemize}
    \item We first \textbf{reveal the misuse risk of LoRA} on model-sharing platforms, raising the alarm for creators and platforms.
    \item We make the first attempt to develop a data-free editing method to secure the sharing of personalized LoRA with adversarial optimization and semantic augmentation.
    \item Extensive and practical experiments demonstrate the effectiveness, robustness, and efficiency of LoRAShield.
\end{itemize}

\section{Background \& Related Work}

\begin{table*}[t]
\tabcolsep=0.2cm
\centering
\caption{Comparison with existing erasing methods. \ding{52}/\ding{56} illustrates whether the method can achieve the corresponding property.}
\vspace{-10pt}
\label{tab:related_work}
\aboverulesep=0.1ex
\belowrulesep=0.3ex
\begin{tabular}{ccccccc}
\toprule
\multirow{2}{*}{\textbf{Erasing Methods}} & \multirow{2}{*}{\textbf{Target}} & \multicolumn{2}{c}{\textbf{Effectiveness}} & \multicolumn{3}{c}{\textbf{Efficiency}} \\
\cmidrule{3-7}
 &  & \textbf{Generalization} & \textbf{Robustness} & \textbf{Data-Free} & \textbf{Time (s)} & \textbf{Memory (GB)} \\ 
\midrule
\midrule
ESD~\cite{gandikota2023erasing} & DMs & \ding{52} & \ding{56} & \ding{52} & $\approx 6100$ & 17.8\\
FMN~\cite{zhang2024forget} & DMs & \ding{56} & \ding{56} & \ding{56} & $\approx350$ & 17.9\\
UCE~\cite{gandikota2024unified} & DMs & \ding{56} & \ding{56} & \ding{52} & $\approx430$ & 5.1\\
EraseDiff~\cite{wu2024erasediff} & DMs & \ding{52} & \ding{56} & \ding{56} & $\approx1500$ & 27.8\\
SPM~\cite{lyu2024one} & DMs & \ding{52} & \ding{56} & \ding{56} & $\approx 29700$ & 6.9\\
\rowcolor{pink}  Ours (LoRAShield) & LoRA & \ding{52} & \ding{52} & \ding{52} & $\approx 14$ & 0.23\\ 
\bottomrule
\end{tabular}
\end{table*}

\subsection{Model-Sharing Platforms}
The proliferation of model-sharing platforms such as Civitai~\cite{civitaimodels}, LiblibAI~\cite{liblib}, and Hugging Face~\cite{huggingface} has democratized access to generative model tools, fostering a ``share and play'' culture where users freely distribute and utilize pre-trained models. However, this accessibility introduces critical risks. While this facilitates innovation, it also creates vulnerabilities: benign users who upload models trained on private data face potential misuse by adversaries. Such risks are exacerbated by the lack of robust safeguards on many platforms; for instance, Civitai's reliance on an ``asymmetric multisided marketplace''~\cite{Griffin24linkin} prioritizes model diversity over rigorous vetting.

\subsection{Concept Erasing in DMs}
\label{sec:concept erasing}
Large-scale DMs are shadowed by their propensity to produce ethically risky or legally contentious outputs, such as sexually explicit imagery, culturally sensitive content, or artistic styles protected by copyright. A growing number of research studies~\cite{li2024safegen,fan2023salun,zhang2024defensive,zhang2024forget,gandikota2023erasing,gandikota2024unified,zhang2024unlearncanvas} have proposed leveraging unlearning techniques to remove or suppress specific concepts via unlearning in generative models to prevent misuse. The traditional machine unlearning strategy~\cite{bourtoule2021machine} aims to modify a model to make it generalize without memorization to tackle privacy and copyright concerns. Although early approaches~\cite{gandikota2023erasing, kumari2023ablating} focused on fine-tuning all DMs parameters to forget a target concept. For example, the Erased Stable Diffusion (ESD) by Gandikota et al.\cite{gandikota2023erasing,zhang2024forget,kumari2023ablating} fine-tunes a DM with a negative guidance teacher to align the given visual concept with `` '' condition. Similarly, Zhang et al.\cite{zhang2024forget} proposed a ``Forget-Me-Not'' (FMN) technique that aligns the model output for the target concept with that of a benign ``anchor'' concept, thus preventing the model from generating the target concept under its text condition, allowing the model to forget without retraining from scratch while preserving closely related concepts. However, since these methods tend to struggle with sizeable computational costs~\cite{gandikota2024unified,zhang2024unlearncanvas}, Unified Concept Editing (UCE)~\cite{gandikota2024unified} proposed to edit DMs without training to improve the efficacy and scalability of the previous work. Wu et al.\cite{wu2024erasediff} formulated diffusion unlearning as a constraint optimization and achieved it by deviating the learnable generative process from the ground-truth denoising procedure, while the assumption requiring all training data is unrealistic. Later, Lyu et al.~\cite{lyu2024one} argued that most of the existing methods failed to preserve the generation of untargeted concepts and proposed an erasing framework (SPM) via one-dimensional adapters to erase multiple concepts. 

\subsection{Adversarial Attacks against DMs}
\label{sec:t2i attack}
The growing adoption of defensive strategies in DMs has catalyzed parallel research into system robustness and security~\cite{wang2024security}. Recent work demonstrates that DMs remain susceptible to generating harmful content even after deploying safety measures, particularly with adversarially crafted inputs. For example, Yang et al.~\cite{yang2024sneakyprompt} crafts semantically preserved yet misleading prompts on surrogate models, enabling transferable attacks that deceive prompt filters. Instead, MMA-Diffusion~\cite{yang2024mma} conducted transferable attacks by crafting adversarial prompts on surrogate models that can deceive the prompt filter while remaining semantically similar to the target concept. Zhang et al.~\cite{zhang2024generate} further exposed the fragility of concept-erasure techniques by proposing UnlearnDiffAtk, an adversarial prompt generation method that resurrects supposedly ``erased'' concepts in fine-tuned DMs. These papers reveal that DMs subjected to rigorous safeguards are also vulnerable to synonym substitutions, in the lack of robustness against adaptive adversaries. 

\subsection{Remarks}
\label{sec:remarks}
Overall, the existing literature indicates the following points: (1) all of the concept-erasing methods target DMs instead of LoRAs, which are frequently used for customization of individuals; (2) recent evaluation~\cite{zhang2024unlearncanvas} also reveals that most of these attacks exhibit limited generalization to in-domain prompts (innocent and not relevant to the erased concept); (3) the majority of these methods fail to defense against current attacks~\cite{yang2024sneakyprompt,zhang2024generate,yang2024mma}; (4) balancing completeness of erasure with minimal data, time and computation resources is still challenging. In comparison, as shown in Tab.~\ref{tab:related_work} where generalization and robustness refer to performance against semantically related but unseen prompts and resilience against dedicated attacks. LoRAShield addresses these gaps and secures the modular components most vulnerable to misuse in platforms.

\section{Preliminaries}
\subsection{Text-to-Image Diffusion Models}
\label{sec:t2i}
T2I models~\cite{dhariwal2021diffusion} can synthesize high-fidelity images $x\in\mathcal{X}$ by iteratively denoising latent representations $z_t\in\mathcal{Z}$ conditioned on textual inputs $c$. The core framework typically comprises four components: an image encoder $\mathcal{E}:\mathcal{X}\rightarrow\mathcal{Z}$ and decoder $\mathcal{D}:\mathcal{Z}\rightarrow\mathcal{X}$ of the autoencoder, a diffusion model (UNet) $\epsilon_{\theta}$, and a text encoder $\mathcal{T}$. The encoder first compresses images into a lower-dimensional latent space and reconstructs them via a decoder $\mathcal{D}$, i.e., $\mathbf{z} = \mathcal{E}(\mathbf{x})$ and $\hat{\mathbf{x}} = \mathcal{D}(\mathbf{z})$. While $\epsilon_{\theta}$ then operates in this latent space, iteratively refining the noisy representations $\mathbf{z}_t$ through a reverse diffusion process:  
\begin{equation}
\mathbf{z}_{t-1} = \frac{1}{\sqrt{\alpha_t}} \left( \mathbf{z}_t - \frac{1-\alpha_t}{\sqrt{1-\bar{\alpha}_t}} \cdot \epsilon_\theta(\mathbf{z}_t, t, \mathbf{c}) \right) + \sigma_t \cdot \epsilon,
\end{equation}
where timestep $t\in\{1,2,\dots, T\}$, $\epsilon\sim\mathcal{N}(0, \mathbf{I})$,  $\epsilon_\theta$ predicts the noise, $\mathbf{c}=\mathcal{T}(\mathcal{C})$ denotes text embeddings given the prompt set $\mathcal{C}$, and $\alpha_t, \bar{\alpha}_t, \sigma_t$ are scheduler hyperparameters. Specifically, the $\epsilon_{\theta}$ integrates textual guidance via cross-attention layers, which dynamically align image features with target concepts. Cross-attention (CA) layer computes:
\begin{equation}
\text{Atten}(\mathbf{Q}, \mathbf{K}, \mathbf{V}) = \text{softmax}\left( \frac{\mathbf{Q}\mathbf{K}^T}{\sqrt{d_k}} \right)\mathbf{V},
\end{equation}
where queries $\mathbf{Q}$ derive from image tokens, and keys $\mathbf{K}$ and values $\mathbf{V}$ derive from $\mathbf{c}$. This mechanism enables fine-grained control over concept representation, as the model prioritizes text-relevant features during denoising. 

\subsection{Low-Rank Adaptation of Diffusion Models}
\label{sec:lora}
The LoRA~\cite{hu2022lora} enable efficient and effective adaptation and customization of T2I models via decomposing the weight update matrix $\Delta W\in\mathbb{R}^{m\times n}$ when fine-tuning the original weight $W_0$:
\begin{equation}
    \Delta W = B\times A
\end{equation}
where $B\in\mathbb{R}^{m\times r}$, $A\in\mathbb{R}^{r\times n}$ and $r\ll\mathrm{min}(m,n)$ stands for the rank that constrains the update to specified dimension space, significantly reducing the trainable parameters of $W_0$. During the inference stage, $W_0$ is updated with $W_0 + \alpha\Delta W$, where $\alpha$ is a scaling factor that controls the strength of the adaptation.

The rise of platforms like Civitai~\cite{civitaimodels} has democratized access to personalized generative models. Unlike full diffusion models (e.g., SD1.5~\cite{sd15} and its extended versions like DreamShaper~\cite{dreamshaper} and Realistic Vision~\cite{realistic}), which are prohibitively large (often more than 5GB) and computationally expensive to share, LoRA models compress adaptation into lightweight matrices (typically about 100MB). Users train LoRAs on local data (e.g., personal artwork or portraits) and upload them with metadata specifying compatible base models. There are much more LoRAs but less base models on Civitai~\cite{civitaimodels}. This modular approach allows others to: (1) Efficiently combine concepts: Load multiple LoRAs (e.g., ``cyberpunk style'' + ``celestial lighting'') into a single base model, scaling their influence via $\alpha$; (2) Preserve reproducibility: Base models act as standardized backbones, ensuring consistent generation across users.

\subsection{Threat Model}
\label{sec:threat model}
\begin{figure}[t]
    \centering
    \includegraphics[width=\linewidth]{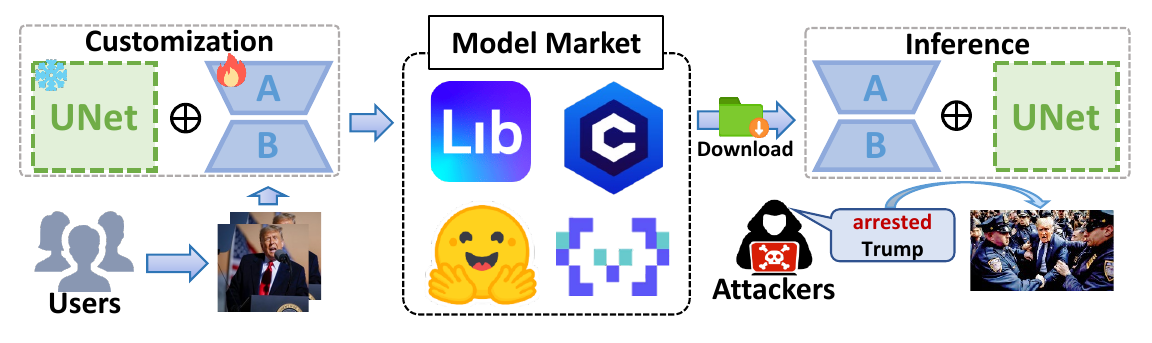}
    \vspace{-10pt}
    \caption{Overview of the threat model.}
    \vspace{-5pt}
    \label{fig:threat_model}
\end{figure}

We consider a scenario involving three parties as depicted in Fig.~\ref{fig:threat_model}: a LoRA owner (victim) who wants to share her/his customized LoRA trained with private data, a trusted model-sharing platform (defender)~\cite{civitaiprotocol} and a LoRA misuse infringer (attacker). 
\subsubsection{LoRA owner's Goal \& Capabilities } The LoRA owner aims to release his/her customized LoRA trained with private images (personal portrait and promoted products) to the public for benign purposes (e.g., personalized photography or advertisement). However, she/he does not want the models to be used for malicious purposes that may affect her/his interests and reputation, e.g., exploiting released LoRA to generate fake images to fabricate rumors about the LoRA owner. Since the owner has no control over the LoRA downloaded by others, to avoid infringements, the owner could request the platform for active editing services to erase the unwanted concept before releasing the LoRA.

\subsubsection{Adversary's Goal \& Capabilities} An attacker on this platform could be anyone who downloads a publicly available LoRA model and attempts to use it for harmful content generation that may violate the platform's content policies~\cite{civitaiprotocol,liblib_protocol}. The generated content~\cite{civitaiprotocol,liblib_protocol} could include violent imagery, sexual or pornographic scenes, deepfakes of real individuals, propaganda, or other disallowed material, leveraging the specific capabilities of the LoRA. Since the attacker can download any LoRAs from the platform, the attacker (1) has complete access and control to the protected LoRA, (2) has complete control over the generation process, but (3) has no access to the training data or unprotected LoRA. However, she/he also wants to ensure the high quality and semantic alignment of the generated images.

\subsubsection{Adversary's Strategies} \label{sec:adversary_strategy} We assume that the attacker is motivated and technically savvy, willing to launch adaptive attacks to bypass the protection. Note that even if most DMs like SD have safety checkers but offline attackers can turn them off or find novel prompts to evade them. (\textbf{Attack I}) One likely strategy is prompt engineering, trying various descriptive prompts (or even gibberish tokens) to trick the model into revealing the protected concept (if the LoRA was meant to be restricted). This aligns with the red-teaming approaches in recent research~\cite{yang2024sneakyprompt, zhang2024generate}, where attackers find that even ``safe'' models can produce unsafe outputs with cleverly crafted prompts. (\textbf{Attack II}) Another powerful strategy is to combine multiple LoRAs since these adapters are often composable, and an adversary might chain a benign LoRA with another one that introduces unsafe elements. For example, they might apply an ``NSFW unlock''~\cite{tools} or violence-enhanced LoRA alongside the protected LoRA, hoping to restore any pruned capabilities or amplify forbidden content. By blending LoRAs, the attacker tries to circumvent a single LoRA's built-in safety, even if our defense cripples the LoRA's ability to generate certain content on its own, the attacker could supplement it with another malicious adapter that fills in the gaps. 
(\textbf{Attack III}) Adversary may adapt LoRA with other base models, while such incompatibility leads to a degradation in generation quality and unpredictable artifacts. This undermines the attacker's goal to create convincing and high-quality harmful imagery. Such constraint is acknowledged by Civitai as well~\cite{civitaiprotocol}, which explicitly warns users that freely combining incompatible model: ``may produce unexpected or lower-quality results''. (\textbf{Attack IV}) Similarly, attackers could also fine-tune (FT) LoRA on malicious data. Considering that the principle of LoRAShield indicates that FT forces a critical trade-off and will corrupt LoRA's benign functionality. Note that there could be more strong attackers which are emphasized in Sec~\ref{sec:limitations}. Based on the above analysis, this paper only consider \textbf{Attack I \& II}. More rationales refer to Appendix~\ref{app:transfer}.

\subsubsection{Defender's Goal \& Capabilities} The defender's (platform) goal is to enable creative uses of LoRA while preventing misuse. For instance, a LoRA might add a new artistic style, and the platform wants regular users to enjoy it but disallows anyone from using it to generate hate symbols or obscene content in that style. Neither the benign users nor the defender trusts every user and thus implements protections specified by the LoRA owner. Therefore, the defender cannot expend significant resources and defense must be automated and efficient (no lengthy retraining or manual intervention for each model). This is because on a large platform like Civitai, there could be thousands of uploaded LoRAs~\cite{wei2024exploring} in a short period, and users expect to upload and update them frequently. Besides, the defender is not accessible to the benign training data of the LoRA for privacy reason, which means that the editing should be data-free as well.
\section{Methodology}
\subsection{Design Overview}
\label{sec:design overview}
Under the threat model presented above, our primary goal is to provide an efficient and robust defense mechanism capable of preventing unauthorized and malicious exploitation of large-scale shared LoRAs on platforms while maintaining their original generative capabilities for legitimate users. In summary, our method addresses several critical challenges below: \textbf{(1) Data-Free:} Since the platform hosting the LoRAs doesn't have access to the original data, the proposed method must operate without reliance on the training dataset. \textbf{(2) Preservation of Utility:} A practical defense should precisely remove only potentially harmful concept representations without diminishing the model’s general utility for creative generation. \textbf{(3) Lightweight and Efficiency:} Given the high frequency and large volume of LoRAs uploaded, our defense must be computationally lightweight to process numerous models quickly without introducing significant latency or computational overhead. \textbf{(4) Robustness Against Adaptive Attacks:} Adversaries are aware of defenses and can employ strategies (Sec \ref{sec:adversary_strategy}) to evade. Hence, the defender must proactively anticipate such behaviors. In the following subsections we overcome these challenges by designing a lightweight framework that effectively balances security needs with usability, raising the barrier to malicious exploitation.


\subsection{Robust Editing Alignment}
As mentioned above in Sec~\ref{sec:t2i}, given the conditional prompt $\mathcal{C}$, the CA layer matters a lot in controlling the image generation. This motivates us to erase the concept in the CA layer to safeguard LoRAs. Initially, we obtain the text embeddings $c\in\mathcal{C}$ and $c_t\in\mathcal{C}_t$ , with the benign and target concepts to be erased, respectively. Further, we calculate the difference between the LoRA-adjusted and original weight matrices for each CA layer. We iteratively update these matrices to minimize the discrepancy between the generated embeddings from the adjusted model and the intended safe embeddings, employing a loss function that combines mean squared error (MSE).
\begin{equation}
\begin{aligned}
    &\mathcal{L}_{align}(c_t, c)=\\&\mathbb{E}_{c_t\sim\mathcal{C}_t, c\sim\mathcal{C}}[\sum_{l\in L}||f^l(c_t; W^l+\alpha\Delta\hat{W}^l)-f^l(c; W^l+\alpha\Delta W^l)||_2^2]
\end{aligned}
    \label{eq:edit align}
\end{equation}
where $\Delta\hat{W}$ means the edited LoRA update matrix. In a CA block, given query $Q$ (image) and key/value $K, V$ (text), the alignment occurs through $K/V$ projections. Specifically, for a CA layer $l$, we will write: $K_t^l=P_K^l(c_t), V_t^l=P_V^l(c_t)$ and $K^l=P_K^l(c), V^l=P_V^l(c)$, where $P_K^l, P_V^l$ denoting the linear projections parameterized by $W^l+\alpha\Delta W^l$ (benign) or $W^l+\alpha\Delta\hat{W}^l$ (edited).
However, minimizing $\mathcal{L}_{align}$ without restricting the difference between the edited and original LoRA parameters disrupts the benign performance of the LoRA as well. Therefore, we add the regularization term to preserve the performance on the untargeted concept.
\begin{equation}
        \mathcal{L}_{pre} = \|\Delta \hat{W}-\Delta W\|_2^{2}
\end{equation}
This targeted pruning ensures minimal disruption to benign concept generation while effectively disabling restricted concepts.

\subsubsection{Adversarial Optimization for Robust Generalization} However, incorporating $\mathcal{L}_{align}$ and $\mathcal{L}_{pre}$ can only prevent the generation given the exact same word of the erased concept, but would fail to generalize to the semantic space. Ideally, our goal is to achieve robust alignment by erasing the semantics of the target concept. To achieve this, we first make the following assumption.

Let $w_{t}^{i}\in\mathcal{W}_t$ be the set of synonyms with the same concept in the text space. We assume that $\forall w_{t}^{i},w_{t}^{j}\in\mathcal{W}_t$ and $\exists\gamma>0$, and we have $\mathrm{Dist}(\mathcal{T}(w_{t}^{i}), \mathcal{T}(w_{t}^{j}))\leq\gamma$, where $\mathrm{Dist}(\cdot)$ measures the difference between the word embeddings~\cite{madry2017towards,radford2021learning}. Note that this assumption is particularly justified in high-quality pretrained models, such as CLIP~\cite{radford2021learning}, where semantic neighbors in the text space are often mapped to proximity in vector space, supporting a smooth and compact representation of conceptual similarity. With the assumption above, using $\ell_2$ to stand for $\mathrm{Dist}(\cdot)$, we formalize the target concept as a $\ell_2$-bounded region in the text embedding space. Specifically, for a target concept with text embedding $c_{t}$, we define its semantic neighborhood as an $\ell_2$ ball: 
\begin{equation}
    \mathcal{B}(c_{t}, \gamma) = \{c_{t}^{\prime} \in \mathbb{R}^d \mid \|c_{t}^{\prime} - c_{t}\|_2 \leq \gamma \},
\end{equation}
where $\gamma$ bounds the maximum distance between $c_{t}^{\prime}$ and any synonym embedding. That is, $\mathcal{L}_{align}$ can be further expressed as:
\begin{equation}
    \mathbb{E}_{\hat{c}_t\in\mathcal{B}(c_{t}, \gamma)}\|\hat{c}_t\times(W + \alpha\Delta \hat{W}) - c\times(W + \alpha\Delta W)\|_2^{2}
\end{equation}
where $\Delta \hat{W}$ refers to the edited LoRA weight matrix optimized with LoRAShield. We notice that the minimization of this expectation can be transformed into a bi-level optimization problem by maximizing the lower bound of the expectation:
\begin{equation}
    \min_{\Delta \hat{W}}\max_{\hat{c}_t\in\mathcal{B}(c_{t}, \gamma)}\|\hat{c}_t\times(W + \alpha\Delta \hat{W}) - c\times(W + \alpha\Delta W)\|_2^{2}.
\end{equation}
However, directly solving this problem often involves large computations since solving the inner maximization is not trivial~\cite{madry2017towards}. Drawing inspiration from recent works~\cite{wu2020adversarial,chen2024enhancing} illustrating that the adversarial optimization in sample space can also be achieved with adversarially robust parameters within the restricted $\ell_{2}$ bound:
\begin{equation}
    \min_{\Delta \hat{W}}\max_{\|\delta_{w}\|\leq\tau} \mathbb{E}\|c_t\times(W + \alpha\Delta \hat{W} + \delta_{w}) - c\times(W + \alpha\Delta W)\|_2^{2}
    \label{eq:minmax}
\end{equation}
Here, $\delta_w$ represents adversarial perturbations to the LoRA parameters, constrained by $\tau$ to balance robustness and stability. Equation~\ref{eq:minmax} shifts the optimization focus from the input embedding space to the parameter space, enabling efficient gradient-based updates while implicitly covering semantic variations within $\mathcal{B}(c_t, \gamma)$. 

\subsubsection{Low-Rank Decomposition via Singular Value Decomposition (SVD)}
\label{sec:svd}
Meanwhile, we should emphasize that directly optimizing $A, B$ to obtain $\Delta \hat{W}$ is not trivial since both matrices are projected into low-dimensional space (non-convexity of matrix factorization~\cite{sun2016guaranteed}), leading to significantly unstable optimization. Instead, we first initialize $\Delta\hat{W} = BA$ and directly update $\Delta\hat{W}$. Subsequently, $\Delta\hat{W}$ is decomposed via SVD. This step is crucial for reducing the optimization complexity while maintaining the structural integrity of the model's learned representations. Given a weight matrix $\Delta\hat{W}$, we decompose it into two lower-rank matrices $\hat{A}$ and $\hat{B}$, which serve as a compact approximation, ensuring effective pruning alignment with minimal performance loss. Formally, we perform SVD on $\Delta\hat{W}$:
\begin{equation}
\Delta\hat{W} = U S V^T, 
\label{eq:svd}
\end{equation}
where $U$ and $V$ are the left and right singular vector matrices, and $S$ is the diagonal matrix containing singular values. We retain only the top $r$ singular values and their corresponding singular vectors, indicated as $U_{r}, S_r, V_r^T$, where $r$ is the rank of LoRA. To construct the final low-rank approximation, we compute the square root of the singular values: $\Sigma_r^{\frac{1}{2}} = \operatorname{diag}(\sqrt{S_r})$
The resulting decomposition is then defined as: $\hat{B} = U_r \Sigma_r^{\frac{1}{2}}, \quad \hat{A} = \Sigma_r^{\frac{1}{2}} V_r^T$
Thus, the original weight matrix $ \Delta\hat{W} $ is approximated as: $\Delta\hat{W} \approx \hat{B}\hat{A}$
Theoretically, we agree that SVD is lossy, but it's controllable~\cite{ZhangCBH0CZ23,MengWZ24}, guaranteed by the Eckart-Young-Mirsky theorem~\cite{YuS11}. Empirically, we found that such an approximation has little influence on the overall performance. Detailed explanation refers to Appendix~\ref{app:svd}.

\begin{algorithm}[t]
\caption{Robust Editing with Semantic Augmentation}
\label{alg:lorashield}
\begin{algorithmic}[1]
\REQUIRE Clean T2I diffusion model with text encoder $\mathcal{T}$ and UNet $\epsilon_{\theta}$; the target concept $\mathcal{C}_t$ to be erased; the target LoRA model with parameter $\Delta W$; total editing step $T$.
\ENSURE Aligned LoRA model.
\STATE Initialize $\Delta \hat{W} \leftarrow\Delta W$
\STATE $\{c_{t}^{1},c_{t}^{2},\dots,c_{t}^{K}\}, \{c^{1},c^{2},\dots,c^{K}\}\leftarrow\mathcal{T}(\mathrm{LLM}(\mathcal{C}_t))$ // Sec~\ref{sec:sa}
\FOR{$t=0,1,\cdots,T-1$}
    \STATE Compute the perturbed $\delta_{w}$ with $\tau\cdot\nabla_{W}\mathcal{L}_{align}$
    \STATE Solving the inner maximization $\hat{W}\leftarrow W + \delta_{w}$
    \STATE // Compute the loss $\mathcal{L}_{all}$ based on $\hat{W}$.
    \STATE $\mathcal{L}_{all}\leftarrow\mathcal{L}_{align} + \eta\cdot\mathcal{L}_{pre}$
    \STATE Update $\Delta\hat{W}$ via $\nabla_{\Delta\hat{W}}\mathcal{L}_{all}$ using Adam.
\ENDFOR
\STATE Approximate $\Delta\hat{W}$ using SVD in Equation~\ref{eq:svd}.
\RETURN $\Delta\hat{W}$
\end{algorithmic}
\end{algorithm}

\subsection{Semantic Augmentation} 
\label{sec:sa}
While adversarial optimization allows the erasure of the embedding space around the given concept, it remains confined to local neighborhoods defined by the $\ell_2$ ball $\mathcal{B}(c_t, \gamma)$ and fails to account for global semantic relationships that attackers may exploit through paraphrasing or cross-concept blending. To address this limitation, we propose a semantic augmentation strategy that expands the defended semantic space by enriching the conceptual representation of the target to be erased. Our method operates without requiring access to the original training data, instead leveraging LLMs to generate semantically relevant synonyms and antonyms for the target concept. Formally, given a target concept $c_t$, we query an LLM to produce: (1) Synonyms: Terms closely aligned with $\mathcal{C}_t$ (e.g., ``modern'' $\rightarrow$ ``contemporary'') to capture linguistic variations; (2) Antonyms: Terms contrasting with $\mathcal{C}_t$ (e.g., ``modern'' $\rightarrow$ ``archaic'') to define conceptual boundaries. For concepts with no viable antonyms, we default to a neutral reference embedding (embedding of ``""). This ensures that the erased concept is suppressed without requiring a contrastive counterpart, preserving alignment stability. These augmented terms are then encoded into text embeddings synonyms $\{c_{t}^{1},c_{t}^{2},c_{t}^{3},\dots,c_{t}^{K}\}$ and antonyms $\{c^{1},c^{2},c^{3},\dots,c^{K}\}$ ($K=5$ in this paper) to replace the $c_{t}$ and $c$ in Equation~\ref{eq:minmax}. By integrating these embeddings, we enforce robustness against a broader range of adversarial prompts. The overall procedure incorporating ``Semantic Augmentation'' is illustrated in Alg~\ref{alg:lorashield} and Appendix~\ref{app:alg}.

\begin{table*}[t]
\centering
\caption{Performance of LoRAShield on three base models and four datasets. Each metric contains two values for edited (1st) and benign (2nd) LoRA, respectively. The $\uparrow$ ($\downarrow$) indicates that a higher (lower) value for the metric signifies superior performance.}
\label{tab:main}
\vspace{-10pt}
\tabcolsep=0.2cm
\renewcommand{\arraystretch}{0.8}
\aboverulesep=0ex
\belowrulesep=0.5ex
\begin{tabular}{ccccccccc}
\toprule
\multirow{2}{*}{\textbf{Base Model}} & \multirow{2}{*}{\textbf{Datasets}} & \multicolumn{2}{c}{\textbf{Effectiveness}} & \multicolumn{3}{c}{\textbf{Functionality-preserving}} & \multicolumn{2}{c}{\textbf{Efficiency}} \\ 
\cmidrule{3-9}
 &  & CRS$\downarrow$ & IRS$\downarrow$ & BPS$\uparrow$ & FID$\downarrow$ & LPIPS$\downarrow$ & Time(s) & Memory(GB)  \\ 
\midrule
\midrule
\multirow{4}{*}{\textbf{SD1.5}} & pixelart & 13.98/17.72 & 11.44/13.22 & 32.06/32.34 & 67.97/94.92 & 0.37/0.67 & 12.97 & 0.2246 \\
& 3DM & 15.24/19.71 & 14.53/17.58 & 30.99/30.56 & 56.44/63.38 & 0.39/0.60 & 13.18 & 0.2363 \\
& trump & 12.70/14.08 & 12.08/13.63 & 25.56/25.27 & 59.60/77.70 & 0.29/0.67 & 13.56 & 0.2246 \\
& Cat & 12.57/14.13 & 12.72/13.42 & 29.56/29.71 & 77.58/69.71 & 0.38/0.62 & 12.2 & 0.2363 \\
\midrule
\multirow{4}{*}{\textbf{DreamShaper}} & pixelart & 12.93/19.30 & 10.68/14.61 & 34.94/35.25 & 70.59/94.34 & 0.41/0.67  & 12.98  & 0.2246  \\
 & 3DM& 15.53/20.43 & 14.73/16.90 & 32.47/32.70 & 40.36/54.88 & 0.30/0.54  & 13.74  & 0.2246  \\
 & trump & 11.34/14.34 & 11.67/14.41 & 26.29/26.35 & 66.21/68.29 & 0.40/0.62  & 17.94  & 0.2178  \\
 & Cat& 12.83/14.62 & 12.50/13.56 & 30.90/31.01 & 68.02/65.21 & 0.37/0.56  & 16.32  & 0.2363  \\ 
\midrule
\multirow{4}{*}{\textbf{Realistic Vision}} & pixelart & 13.20/20.41 & 10.88/14.56 & 35.34/35.01 & 65.27/108.55  & 0.37/0.67  & 13.52  & 0.2246  \\
 & 3DM & 16.55/20.38 & 14.73/17.35 & 31.89/32.44 & 46.50/58.88 & 0.32/0.55  & 11.68  & 0.2363  \\
 & trump & 12.40/14.81 & 12.46/13.56 & 26.22/26.78 & 62.00/69.69 & 0.35/0.61  & 17.05  & 0.2168  \\
 & Cat& 12.75/14.40 & 12.05/13.30 & 31.21/30.84 & 66.62/65.69 & 0.35/0.57  & 13.75  & 0.2207  \\ 
\bottomrule
\end{tabular}
\end{table*}

\section{Evaluation}
\label{sec:evaluation}
\subsection{Experimental Setup}
\subsubsection{Base Models} We select SD v1.5~\cite{sd15} DreamShaper~\cite{dreamshaper} and Realistic Vision~\cite{realistic} as the base models in the experiments, since SD v1.5 is the most popular base model on Civitai and there are many remarkable variants, among which DreamShaper and Realistic Vision, with 1.2 and 1.6 million downloads on Civitai alone~\cite{civitaimodels}. 

\subsubsection{Datasets and Personalization Methods}
We select four datasets for the evaluation in this paper, including two style-based datasets: (1)``3DM''~\cite{3DM} is a dataset that has 3D rendering style;(2)``pixelart''~\cite{pixelart} is a dataset that has pixel-like style; and two portrait-based datasets (3)``Cat''~\cite{cat} that describe a cat with all kinds of dressings; and (4)``trump''~\cite{trump} that has the images of Donald Trump generated by LoRA. For LoRA fine-tuning, we set the rank at 128, targeting the attention and projection layer of the UNet with 500 steps (lr=2e-5) following the code given by Hugging Face~\cite{huggingface25lora}. The implementation details of LoRAShield are presented in Appendix~\ref{app:impl}

\begin{figure}[t]
    \tabcolsep=0.05cm
    \centering
    \begin{tabular}{cc|c}
    \toprule
    \textbf{Prompt} 
    & \textcolor{blue}{Cat},\textcolor{blue}{$\dots$} 
    & \textcolor{blue}{Cat}, \textcolor{red}{bloody}\textcolor{blue}{$\dots$} \\
    \midrule
    \midrule
    \raisebox{2.5\height}{\makecell{\textbf{Benign}\\ \textbf{LoRA}}}
    &\raisebox{0\height}{\includegraphics[width=0.25\linewidth]{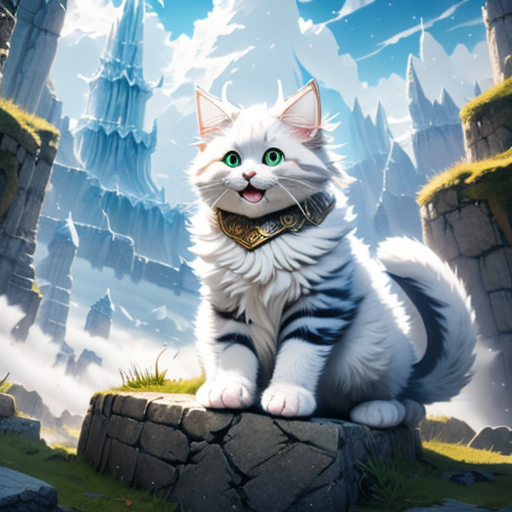}} 
    & \raisebox{0\height}{\includegraphics[width=0.25\linewidth]{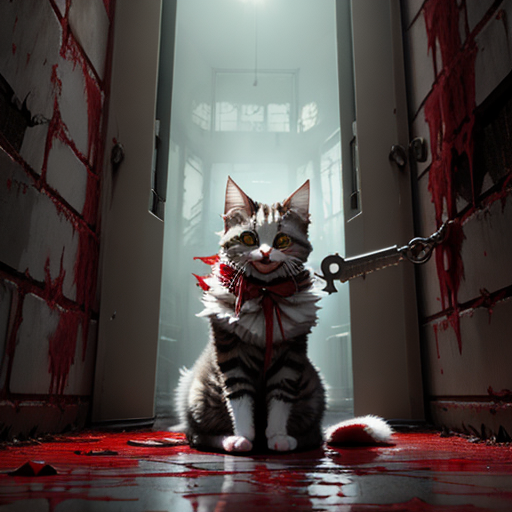}} \\
    \midrule
    \raisebox{2.5\height}{\makecell{\textbf{Edited}\\ \textbf{LoRA}}}
    &\raisebox{0\height}{\includegraphics[width=0.25\linewidth]{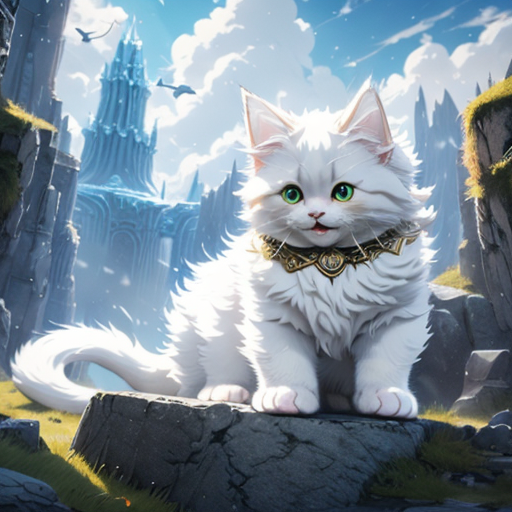}} 
    & \raisebox{0\height}{\includegraphics[width=0.25\linewidth]{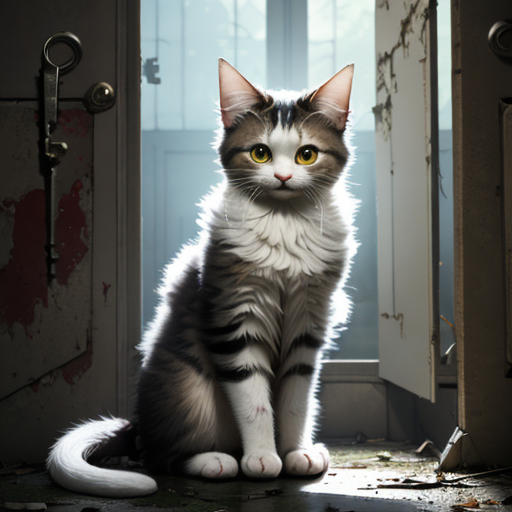}} \\
    \bottomrule
    \end{tabular}
    \vspace{-10pt}
    \caption{Images synthesized with benign (1st row) and edited (2nd row) LoRA when the given prompts \textcolor{blue}{without} (1st column) and \textcolor{red}{with} (last column) target concept.}
    \label{fig:figure}
\end{figure}

\subsubsection{Evaluation Metrics} Following the previous works~\cite{zhangunlearncanvas, zhang2024defensive,gandikota2024unified}, the effectiveness of LoRAShield is assessed with the following metrics. (1) Concept Removal Score (CRS) measures the similarity between images generated with prompts with the given word and the target concept using the CLIP score~\cite{radford2021learning}, and a lower CRS denotes better erasing performance. Similarly, (2) In-domain Retain Score (IRS) measures the similarity between images generated with prompts of the target concept but different from the given word and the target concept using the CLIP score. (3) Benign Preservation Score (BPS) measures the similarity between the images and the prompts without the undesired concept to evaluate the faithfulness of the image and corresponding prompt using the CLIP score. (4) FID~\cite{heusel2017gans} compares the quality and fidelity between the images generated with benign LoRA and images generated with edited LoRA. A lower score means that the distribution of images is more similar. (5)LPIPS compares the semantic similarity between the image generated on benign and the edited LoRA with benign prompts. In particular, for nudity measurement on ``3DM'' and ``pixelart'' datasets, following the previous work~\cite{li2024safegen}, we use NudeNet~\cite{NudeNet} to further validate the effectiveness of LoRAShield.  The efficiency is measured by running the (1) Time of the erasing process and the (2) Memory costs to conduct the erasing. The robustness of LoRAShield against out-of-domain prompts and attacks is presented in Sec~\ref{sec:robustness}.

\subsection{Performance against Misuse} 
The performance is presented in Tab.~\ref{tab:main}. Note that for a fair comparison, we also report the performance of benign LoRA. To measure the influence of LoRAShield on FID and LPIPS, we regenerate 500 images with different random seeds following the setting before as a comparison for benign LoRAs.

\begin{table}[h]
\centering
\caption{Nudity of images generated with edited and benign LoRA on three base models and two datasets. }
\label{tab:nudity}
\vspace{-10pt}
\tabcolsep=0.2cm
\renewcommand{\arraystretch}{0.8}
\aboverulesep=0ex
\belowrulesep=0.5ex
\begin{tabular}{cccc}
\toprule
\textbf{Base Model} & \textbf{Dataset} & \textbf{Edited} & \textbf{Benign} \\
\midrule
\midrule
SD1.5 & 3DM & 0.09 & 0.67 \\
SD1.5 & pixelart & 0.06 & 0.28 \\
DreamShaper & 3DM & 0.02 & 0.63 \\
DreamShaper & pixelart & 0.09 & 0.55 \\
Realistic & 3DM & 0.13 & 0.74 \\
Realistic & pixelart & 0.09 & 0.59 \\
\bottomrule
\end{tabular}
\end{table}

\subsubsection{Erasing Effectiveness} Tab.~\ref{tab:main} illustrates that the images generated by edited LoRA exhibit much lower CLIP scores compared to the benign one, indicating that LoRAShield can erase the target concept effectively. Additionally, when facing in-domain prompts, the edited LoRA can still maintain a low IRS value. Intuitively, we visualize the generated images (on DreamShaper) in Fig.~\ref{fig:figure}, where the images generated on benign LoRA can generate images that violate the protocol of model-sharing platforms. In contrast, even given the explicit prompts, the images generated on edited LoRA present no target concept while aligning with the other semantic prompts. Furthermore, we evaluate the performance of LoRAShield to erase nudity with NudeNet~\cite{NudeNet} in Tab.~\ref{tab:nudity}, in which the images generated on edited LoRA exhibit a significant decrease in nudity and our in-person check further found that many images generated on edited LoRA that were noted as nude by NudeNet, have been misclassified. More visualizations refer to Appendix~\ref{app:exp}.
\begin{figure}
    \centering
    \includegraphics[width=\linewidth]{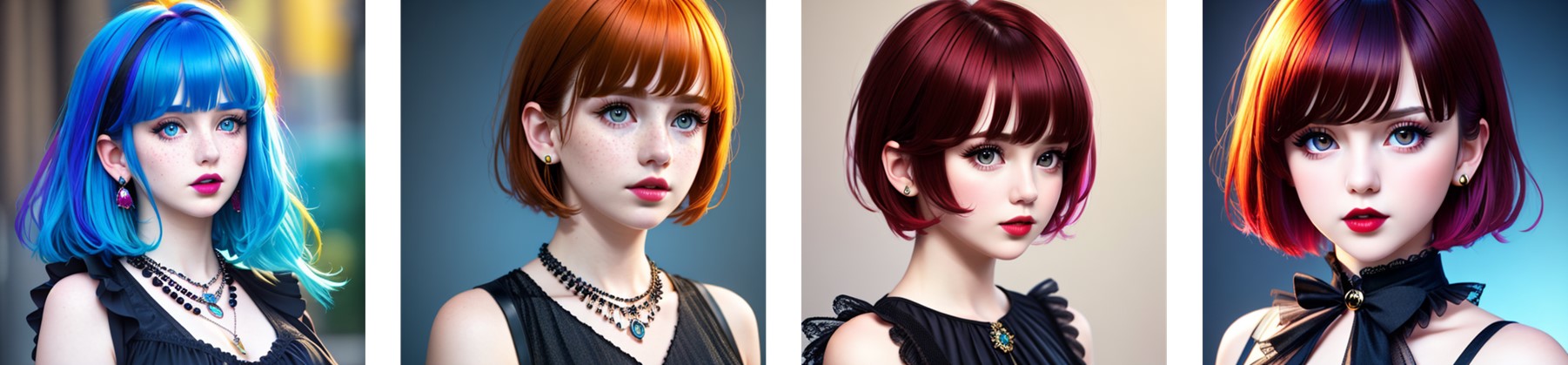}
    \vspace{-10pt}
    \caption{Cumulative concept erasure: From left to right, images generated by (1) benign LoRA, (2) with concept ``hair color'' erased, (3) with concepts ``hair color''+``necklace'' erased, and (4) with concepts ``hair color''+``necklace''+``head gesture'' erased, illustrating progressive suppression of target semantics while preserving non-target features.}
    \label{fig:multiconcept}
\end{figure}

\subsubsection{Functionality-preserving} Other than effectiveness, preserving the functionality of LoRA itself is also necessary. As shown in Tab.~\ref{tab:main}, the BPS values of edited LoRA have a slight decrease in some cases, but most keep close to that of benign LORA. This can be attributed to the interrelationship of concepts, and erasing the target concept may affect the representation of another one. Surprisingly, the images generated on edited LoRA have lower FID and LPIPS values than images generated on benign ones (with different random seeds), which also validates that the negative impact of LoRAShield is acceptable, since its influence is much less than the change of random seeds. From the visual comparison of the images generated on the edited and benign LoRAs in Fig.~\ref{fig:figure}, we can conclude that LoRAShield can effectively preserve the core functionality of LoRAs, aligning with our design goal of balancing robustness with minimal disruption to legitimate generation.

\subsubsection{Erasing Efficiency} LoRAShield achieves platform-scale efficiency, which is critical for deployment in model-sharing ecosystems. As shown in Tab.~\ref{tab:main}, editing a LoRA takes 14 seconds (vs. often several minutes for retraining from scratch) with 0.23GB GPU memory. This efficiency stems from our sequential optimization of attention matrices, which avoids full model retraining and leverages LoRA's low-rank structure. Even for platforms hosting 100k+ LoRAs, LoRAShield enables fast processing on a single A100 GPU, ensuring real-time user experiences even during peak uploads.

\subsubsection{Multiple Concept Erasure} To evaluate LoRAShield's ability to handle cumulative semantic suppression, we conducted a dynamic visualization experiment erasing up to three interdependent concepts in Fig.~\ref{fig:multiconcept}, since quantitative metrics (e.g., CLIP scores) become less reliable for multi-concept scenarios.

\subsection{Effectiveness against Adversarial Attack}
\label{sec:robustness}
As mentioned before in Sec~\ref{sec:adversary_strategy}, since malicious users can download the edited LoRA from the platforms, it is critical to validate whether LoRAShield can resist potential attacks. In this paper, we consider four input-based attacks~\cite{yang2024sneakyprompt,yang2024mma,qu2023unsafe,schramowski2023safe} that construct adversarial prompts. Another common scenario is that malicious users may merge multiple LoRAs for misuse, which can disrupt the protection of LoRAShield. For simplicity, we consider ``nude'' as the target concept on ``3DM'' datasets and use ``DreamShaper'' as the base model. More rationales for this are in Appendix~\ref{app:model}.
\begin{table}[h]
\centering
\caption{Performance against multiple LoRA merge.}
\label{tab:fusion}
\vspace{-10pt}
\tabcolsep=0.2cm
\renewcommand{\arraystretch}{0.8}
\aboverulesep=0ex
\belowrulesep=0.5ex
\begin{tabular}{ccccc}
\toprule
\multirow{2}{*}{\textbf{LoRA}} & \multicolumn{2}{c}{\textbf{Benign}} & \multicolumn{2}{c}{\textbf{Edited}} \\ 
\cmidrule{2-5} 
& CRS & Nudity & CRS & Nudity \\ 
\midrule
\midrule
Base & 20.43 & 0.63 & 15.53 & 0.02 \\
\midrule
+Tools~\cite{tools} & 20.60 & 0.70 & 15.36 & 0.08 \\
+Background~\cite{background} & 16.39 & 0.22 & 14.11 & 0.02 \\
+Clothes~\cite{clothes} & 15.11 & 0.10 & 13.77 & 0.06 \\
+Pose~\cite{pose} & 14.99 & 0.06 & 13.64 & 0.00 \\
+Celebrity~\cite{celebrity} & 13.81 & 0.06 & 12.72 & 0.00 \\ 
\bottomrule
\end{tabular}
\end{table}

\subsubsection{Robustness against Multi-LoRA Merge}
A common but critical threat arises when attackers merge multiple edited LoRAs. For instance, combining a ``beach scenery'' LoRA (with ``nude'' erased) and ``human portrait'' LoRA (without ``nude'' erased) might reintroduce NSFW content. We use weight averaging to linearly combine multiple LoRA matrices $\Delta W_{merged} = \sum\alpha_{i}\Delta W_{i}$. Five popular LoRAs of different types from Civitai: tools~\cite{tools} (for adjusting the clothes of people), background~\cite{background}, clothes~\cite{clothes}, pose~\cite{pose} and celebrity~\cite{celebrity} are selected for evaluation with base LoRA trained on ``3DM'' dataset and DreamShaper with $\alpha_{i}=1$. The result in Tab.~\ref{tab:fusion} showcases that even under the perturbation in parameter space, the edited LoRA retains stable performance compared with the base one.

\begin{figure}[h]
    \centering
    \includegraphics[width=0.8\linewidth]{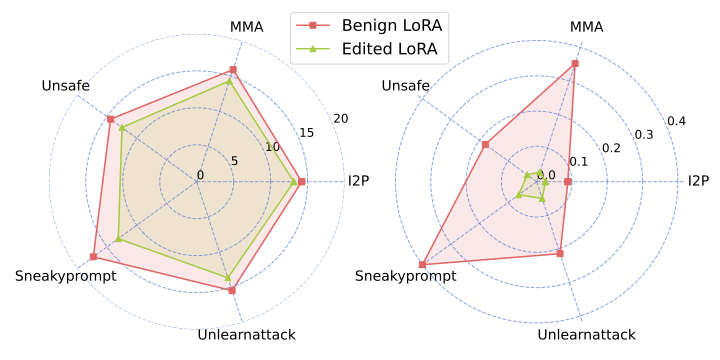}
    \vspace{-10pt}
    \caption{CRS (left) and Nudity (right) under input attacks.}
    \label{fig:attack}
\end{figure}

\subsubsection{Robustness against Input-based Attack}
Malicious users may iteratively refine the prompts to bypass concept erasure, escalating from simple keyword swaps to sophisticated semantic manipulations. To simulate this arms race, we evaluate LoRAShield against four SOTA input-based attacks ~\cite{yang2024sneakyprompt,yang2024mma,qu2023unsafe,schramowski2023safe} that reflect realistic evasion tactics. The result in Tab.~\ref{fig:attack} demonstrates that even under input-based attacks, LoRAShield exhibits remarkable performance. Specifically, the nudity score of images generated by edited LoRA is much lower than that of benign one.

\subsection{Real-world Case Study}
To validate the practical efficacy of LoRAShield, we conducted end-to-end experiments on Civitai. Our threat scenario mirrors real-world misuse: benign LoRAs~\cite{yitong} uploaded by legitimate users are exploited by attackers to generate policy-violating content. For safety consideration, we do not provide the explicit images since it involves adult-oriented content. We can notice that the benign LoRA flagged for generating celebrity portraits but could produce NSFW content when given specified prompt. To mitigate this, we download it and attempt to generate NSFW images with the same prompt on edited LoRA. The result highlights that LoRAShield empowers platforms to proactively neutralize misuse risks without sacrificing creative utility or scalability.

\subsection{Ablation Study}
\label{sec:ablation}
To dissect LoRAShield’s design choices, we performed ablation experiments below. All experiments in this subsection were conducted on the ``3DM'' dataset and DreamShaper.

\subsubsection{Sensitivity Analysis} 
We conduct a detailed sensitivity analysis on the key hyperparameters, namely the number of augmentation $K$ and the threshold $\tau$. As summarized in Tab.~\ref{tab:sensitivity}, the performance measured by CRS and Nudity scores, remains stable across a wide range of values. These results demonstrate that LoRAShield is not sensitive to hyperparameter selection within reasonable ranges, ensuring its practical deployment in diverse scenarios. Also, we tested multiple LLMs for synonym/antonym augmentation with result in Tab.~\ref{tab:llm_dependency}. As shown in the table, LoRAShield remains effective across LLMs, denoting its the robustness and generalization.

\subsubsection{Adaptation to DiT-based Models.} 
To validate our method's extensibility, we adapted LoRAShield to SD3 by focusing on the attention layers within the MM-DiT blocks. Unlike our previous strategy that targeted specific attention weights, this objective allows for the optimization of \textbf{any} LoRA parameters within the MM-DiT block using randomly initialized latents $x_t$. As shown in Tab.~\ref{tab:sd3_results}, the experimental results across various datasets confirm that LoRAShield remains effective on DiT architectures, maintaining low CRS/IRS while preserving competitive image quality.

\begin{table}[h]
\centering
\caption{Ablation study on robust optimization (RO) and semantic augmentation (SA). \ding{52} and \ding{56} denote whether the corresponding techniques are incorporated in LoRAShield.}
\label{tab:ablation}
\vspace{-10pt}
\tabcolsep=0.2cm
\renewcommand{\arraystretch}{0.8}
\aboverulesep=0ex
\belowrulesep=0.5ex
\begin{tabular}{cccc}
\toprule
\textbf{RO} & \textbf{SA} & \textbf{CRS} & \textbf{Nudity} \\ 
\midrule
\ding{56} & \ding{56} & 14.60 & 0.02 \\
\ding{56} & \ding{52} & 14.55 & 0.00 \\
\ding{52} & \ding{56} & 14.53 & 0.00 \\
\ding{52} & \ding{52} & 14.20 & 0.00 \\ 
\bottomrule
\end{tabular}
\end{table}

\subsubsection{Impact of Robust Optimization and Semantic Augmentation}
Since we have claimed that the proposed robust optimization and semantic augmentation enhance the generalization of concept erasure, we validate this with an ablation study. The result in Tab.~\ref{tab:ablation} illustrates that both enhancements can decrease the nudity score of the generated images while incorporating together further improves CRS, validating our design of LoRAShield.

\subsubsection{Impact of Merging Ratio}
\begin{figure}[h]
    \centering
    \includegraphics[width=0.8\linewidth]{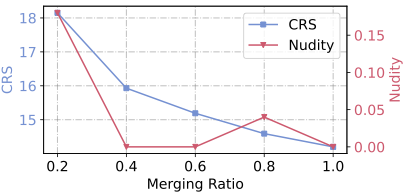}
    \vspace{-10pt}
    \caption{Impact of Merging Ratio.}
    \label{fig:merge}
\end{figure}

The merge $\alpha$ ratio governs the strength of LoRA weights when merging with the base model, and a lower strength may weaken the protection of LoRAShield as well as the benign concept. Here, we set $\alpha$ at \{0.2, 0.4, 0.6, 0.8, 1.0\} to quantify trade-offs between faithfulness and protection and the result in Fig.~\ref{fig:merge} demonstrates that even the nudity score fluctuates slightly but remains at a low level when $\alpha\ge 0.4$. Importantly, we should also emphasize that a low merging ratio will definitely degrade the performance of benign tasks. In real-world misuse scenarios, attackers are incentivized to maintain high $\alpha$ values to preserve the LoRA's utility for legitimate tasks while attempting to bypass safeguards. This constraint ensures that the protection remains effective when adversaries prioritize functional performance.

\section{Limitations}
\label{sec:limitations}
We must acknowledge that our method cannot completely prevent LoRA misuse, since a strong adversary with extra computation resources could train a new LoRA from scratch using the target LoRA-generated data. Instead, our goal is \textbf{to significantly raise the cost and technical barrier for attackers, making casual or large-scale misuse impractical}. For instance, this situation is analogous to the safety challenges in LLMs. Despite continuous efforts in safety alignment and red-teaming, \textbf{it is theoretically impossible to completely prevent all novel ``jailbreaks''}. However, this does not mean we should abandon defensive efforts. On the contrary, we must be aware of the potential risks and continuously build corresponding mitigation measures to address possible threats. Similarly, for the LoRA ecosystem, we believe our major contribution is being the first to identify and reveal this critical vulnerability. Before a perfect solution is devised, the community must first recognize the problem's existence and severity.

\section{Conclusion}
In this paper, we identify critical vulnerabilities in model-sharing ecosystems. While DMs have been extensively studied for concept erasure, LoRA's misuse risk has been largely overlooked, leaving platforms like Civitai exposed to misuse that harms creators' rights and platform integrity. This work bridges this gap by proposing LoRAShield, the first data-free and editing framework to secure LoRAs against adversarial exploitation. By editing LoRA's weight subspace, our platform-driven approach achieves effectiveness and efficiency in suppressing harmful content while preserving functionality for legitimate uses, enabling scalable deployment across millions of shared LoRAs.

\section*{Acknowledgment}
This work was partly supported by the New Generation Artificial Intelligence-National Science and Technology Major Project under No. 2025ZD0123503, NSFC under No. U2441239, U24A20336, and No. 62502433, the China Postdoctoral Science Foundation under No. 2024M762829, 2025M781523 and 2025M781522, Zhejiang Key Laboratory of Decision Intelligence under No. 2025E10006, Zhejiang Provincial Natural Science Foundation Exploration of China under No. LMS26F020003, State Key Laboratory of Cryptography and Digital Economy Security under No. KFYB2504, the Zhejiang Provincial Natural Science Foundation under No. LD24F020002, and the ``Pioneer, Leading Goose'' R\&D Program of Zhejiang under No. 2025C02033 and 2025C01082, NSFC under No. 62502432, NSFC under No. 62402418 and the Ningbo Yongjiang Talent Project.

\bibliographystyle{ACM-Reference-Format}
\bibliography{main}

\appendix
\section{Appendix}
\subsection{Details of Evaluation}
In this supplementary material, we provide content that was not included in our main paper due to space limitations. This includes additional details on implementation, analysis, and additional quantitative results.

\subsubsection{Implementation Details}
\label{app:impl}
For target concept, following previous settings~\cite{li2024safegen,gandikota2023erasing,zhang2024generate}, we select ``nude'' and as the concept to be erased for style-based datasets (i.e., ``3DM'' and ``pixelart''), and ``bloody'' for portrait-based datasets (i.e., ``cat'' and ``trump''). For both benign and edited LoRA, we use LLM (Qwen-Plus~\cite{yang2024qwen2}) to generate 50 prompts with the corresponding trigger word and generate 10 images for each prompt with 10 different random seeds with 500 images for each model and dataset in total. All of the above metrics are evaluated with these generated images. The editing step is set at 10 and $\tau$=1e-5. We set the merging ratio $\alpha$ of LoRA at 1 when merging with the base model by default. For hyperparameters, $\gamma$ represents the semantic neighborhood of a target concept and was set as a fixed and empirical value for simplicity. 

Since Eq.~\ref{eq:edit align} denotes the expectation over the distribution of benign and target concepts (actually, we sample multiple sentences to capture the precise feature of $c$ and $c_t$) and we will clarify it with a more detailed explanation and definition.

\subsubsection{Rationale of Model Selection}
\label{app:model}
Since our goal is to protect the shared LoRA, we do not make any assumption about the base model. Actually, we found that the mainstream base models on Civitai can also generate NSFW images, while LoRAShield succeeds in protecting.

\subsubsection{Rationale of Metrics Selection}
\label{app:metrics}
The CLIP score might be insufficient for a comprehensive evaluation. That is why we employ a suite of metrics for assessment, e.g., FID and LPIPS for utility, and NudeNet for a direct measure of NSFW content. While the improvement on the CRS might seem modest, this is because the CLIP-based metric is not sensitive to the local concept changing (since LoRAShield won't change the visualization of other concepts), but the improvement of the nudity score also validates LoRAShield's effectiveness.

\subsubsection{Datasets and External LoRAs}
\label{app:data}
We have mentioned that each dataset is evaluated with 50 prompts and all of these prompts were generated by LLMs~\cite{yang2024qwen2}. Since we have evaluated the effectiveness of LoRAShield against malicious input-based attacks, we provide details about how the experiments were conducted.
\begin{itemize}
    \item Sneakyprompt~\cite{yang2024sneakyprompt} is a black-box and automated attack framework to jailbreak T2I generative models to generate NSFW images, utilizing reinforcement learning to guide the perturbation of tokens. Considering that this optimization is a time-consuming process, we leverage the prompts provided by Sneakyprompt~\cite{SneakyPrompt-data}, including 200 prompts.
    \item MMA~\cite{yang2024mma} crafts an adversarial prompt that can deceive the prompt filter while remaining semantically similar to the target. We use the generated adversarial prompt~\cite{MMA-data}, which has been shown to be transferable under a black-box scenario.
    \item I2P~\cite{schramowski2023safe} comprises 4,703 inappropriate prompts, encompassing hate, harassment, violence, self-harm, nudity content, shocking images, and illegal activity. Specifically, we selected the prompt with nude concept (931 in total) for evaluation. Note that these prompts are real-user inputs sourced from an image generation website.
    \item Unsafe~\cite{qu2023unsafe} provides 30 manually crafted malicious prompts~\cite{unsafe-data} that describe sexual and bloody content.
    \item UnlearnDiffAtk~\cite{zhang2024generate} reveals the vulnerability of concept-erased DMs with adversarial prompts. We utilize the provided prompts~\cite{UnlearnDiffAtk-data} with nude concept (28 in total) for evaluation.
\end{itemize}
For each of the datasets above, we generate 1 image for each to calculate the corresponding ``Nudity'' and ``CRS''. Additionally, we also downloaded the online LoRAs to merge with the edited ones for evaluation, and the details are given below.

\begin{itemize}
    \item Tool~\cite{tools} (LEOSAM's Clothing +/- Adjuster) is a functional LoRA for adjusting the amount of clothing on the drawn objects. By adjusting the LoRA weight from -1.0 to 1.0, a gradual reduction of the clothing on the drawn objects can be achieved.
    \item Background~\cite{background} (GHIBLI\_Background) provides generated images with beautiful Ghibli scenes that have a powerful expression for both architecture and landscape as a background.
    \item Clothes~\cite{clothes} (hanfu) is a high-quality hanfu lora model to convert the costumes of the characters into Hanfu.
    \item Pose~\cite{pose} (Shirt Tug Pose) offers shirt tug pose for the characters.
    \item Celebrity~\cite{celebrity} (MIMI) is a portrait LoRA of the celebrity Mi Yang in China.
\end{itemize}
Note that all the LoRAs downloaded were not for commercial use, and the generated images will not be posted online to protect intellectual property.

\subsubsection{More Experimental Results}
\label{app:exp}
\begin{figure}[t]
    \centering
    \includegraphics[width=0.8\linewidth]{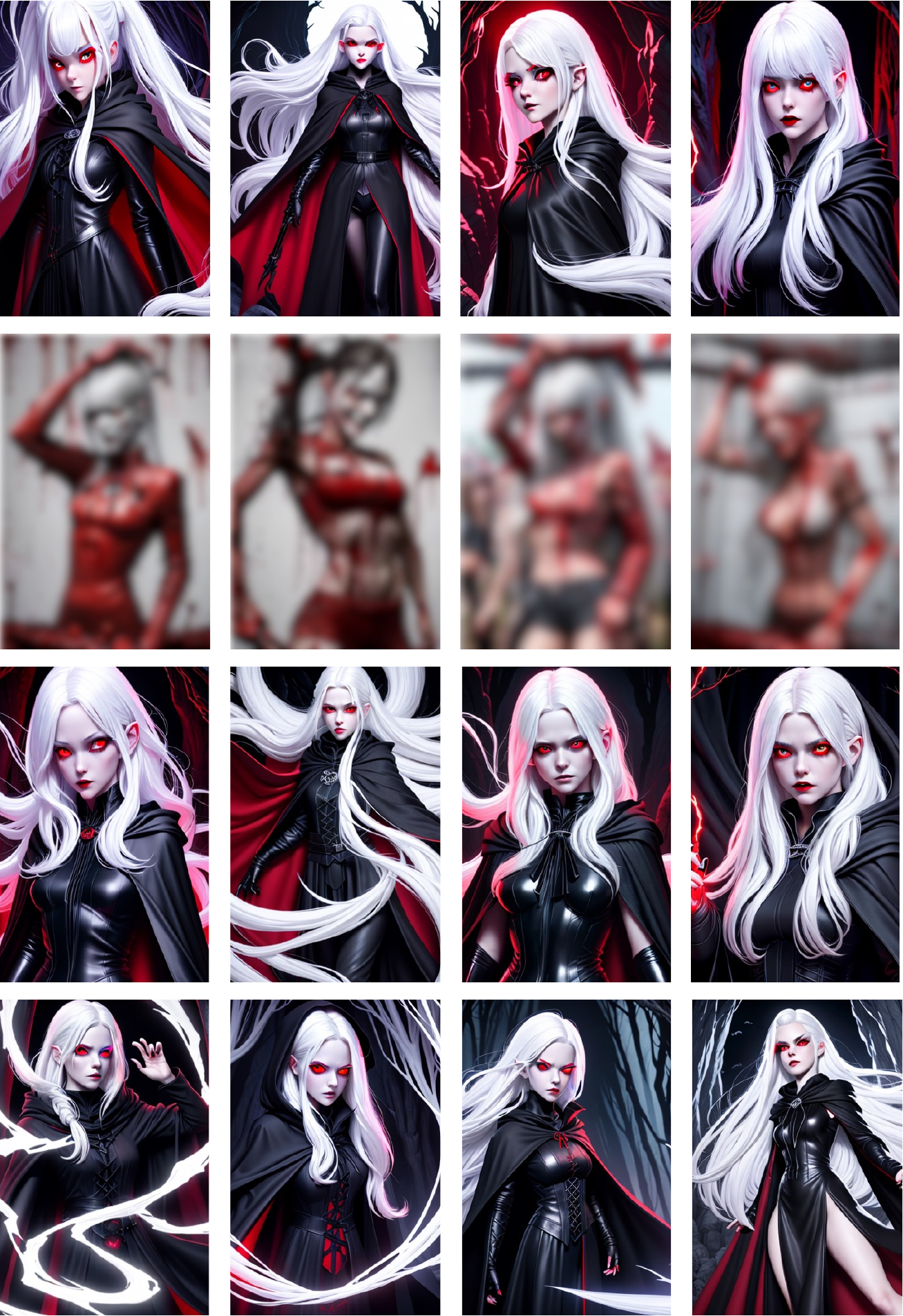}
    \caption{More visualization of concept erasure (``bloody'') on ``3DM'' dataset and DreamShaper. The four rows of images from top to bottom indicate (1) the images generated with benign LoRA on benign prompt; (2) the images generated with benign LoRA with ``bloody'' concept; (3) the images generated with edited LoRA on benign prompt; (4) the images generated with edited LoRA with ``bloody'' concept.}
    \label{fig:ds_3dm}
\end{figure}

For the limited space of the main context, we only provided a little visualization of the generated images by benign and edited LoRAs. Here we provide more results in Fig.~\ref{fig:ds_3dm} to demonstrate LoRAShield's consistency across diverse prompts and base models. These additional results highlight two key observations: (1) edited LoRAs maintain high fidelity to benign prompts (rows 1 vs. 3), and (2) suppression of the ``bloody'' concept remains robust even when explicitly prompted (rows 2 vs. 4), with edited outputs adhering to platform safety guidelines.

\subsection{Rationales of Method and Threat Models}
\label{app:rationale}

\subsubsection{Rationale of LoRA Transfer Attack}
\label{app:transfer}
As mentioned in our paper, LoRAs are finetuned on specific base models (e.g., SDv1.5). Using a LoRA with an incompatible base model leads to a degradation in generation quality and unpredictable artifacts. This undermines the attacker's ability to create convincing, harmful imagery. Our threat model explicitly assumes that the attacker ``wants to ensure the high quality and semantic alignment of the generated images'', which is a realistic assumption for impactful misuse. Such constraint is acknowledged by Civitai as well, which explicitly warns users that freely combining and incompatible model: ``may produce unexpected or lower-quality results. ... Results may vary, and quality is not guaranteed.''

\subsubsection{Technical soundness of the SVD}
\label{app:svd}
Based on the non-convexity of matrix factorization, LoRA leads to unstable gradients, and an analysis is given, with an ablation study vs direct optimization. We agree that SVD is lossy, but it's controllable (supported by AdaLoRA and PissaLora), guaranteed by the Eckart-Young-Mirsky theorem.

\subsubsection{Mitigation for the Semantic Representation}
\label{app:asp}
We admit that use the previous conclusion for semantics representation are not sound, and that is precisely why we employ a two-pronged strategy later:
\begin{itemize}
    \item Robust optimization provides local robustness within the $\ell_2$-ball by training with worst cases in the parameter space.
    \item Semantic augmentation provides global robustness by using an LLM to expand the semantic space beyond the local.
\end{itemize}

\subsubsection{Bi-level optimization}
\label{app:opt}
Since the goal in Eq.7 is to minimize the expectation of $\mathcal{L}_{align}$ over the distribution of $\hat{c}_t$, therefore, we can instead minimize the upper bound (worst-case) of this expectation to make edited LoRA robust against this worst-case attack (prompt-based).

\subsubsection{Correspondence of Alg.1}
\label{app:alg}
The lines in Alg.1 directly map to the sections in our methodology as follows:
\begin{itemize}
    \item Line 2, which uses an $\mathop{LLM}(C_t)$, implements the Semantic Augmentation strategy from Sec 4.3.
    \item Lines 4-8 implement the Robust Editing Alignment procedure from Sec 4.2, including adversarial optimization (Lines 4-5) and combined loss (Line 7). Note that the expression $\hat{c}_t\times (W + \alpha\Delta \hat{W})$ is a simplification representing the interaction within the DM's cross-attention layers, where text embeddings $c$ (as K and V) are combined with image features. 
    \item Line 10 corresponds to the SVD decomposition described in Sec~\ref{sec:svd}.
\end{itemize}

\subsubsection{Adaptation to DiT-based Models}
To validate our adaptation as well, we have tried to transfer LoRAShield to SD3. Specifically, we focus on the attention layer (for LoRA edit alignment) of MM-DiT block in SD3. Given output of 3 text encoder (CLIP-G, CLIP-L, T5), we align the output with: 
$\mathcal{L}_{align}=\mathrm{E}_{c_t\sim\mathcal{C}_t, c\sim\mathcal{C}}[\sum_{l\in L}||f^l(c_t, x_t; \theta^{l})-f^l(c, x_t; \hat{\theta}^l)||_2^2]$, where we use $\theta^l$ and $\hat{\theta}^l$ to represent the parameters (LoRA of joint attnetion weight for SD3) of benign and edited LoRA of layer $l$. Any LoRA paramters in MM-DiT block can be optimized using this objective (with random initiated latent $x_t$), which is different with our previous strategy that only target specific attention weight. This version can be adapted to any model regardless of structures. The experimental results (edited LoRA) are given in Tab.~\ref{tab:sd3_results}. While we should also admit that this version also lead to higher GPU memory cost nealy 1GB with 30s time cost.

\begin{table}[htbp]
\centering
\caption{Results across different LLMs for augmentation.}
\label{tab:llm_dependency}
\small
\setlength{\tabcolsep}{5pt}
\vspace{-10pt}
\tabcolsep=0.2cm
\renewcommand{\arraystretch}{0.8}
\aboverulesep=0ex
\belowrulesep=0.5ex
\begin{tabular}{@{}lcc@{}}
\toprule
\textbf{Backend LLM} & \textbf{CRS} $\downarrow$ & \textbf{Nudity} $\downarrow$ \\ \midrule
GPT-4o               & 14.11          & 0.01               \\
GLM-4.6              & 13.23          & 0.00               \\
Gemini 2.5 Pro       & 14.53          & 0.02               \\
DeepSeek V3          & 15.52          & 0.01               \\ \bottomrule
\end{tabular}
\end{table}
\begin{table}[htbp]
\centering
\caption{Sensitivity analysis of hyperparameters $K$ and $\tau$.}
\label{tab:sensitivity}
\vspace{-10pt}
\tabcolsep=0.05cm
\renewcommand{\arraystretch}{0.8}
\aboverulesep=0ex
\belowrulesep=0.5ex
\begin{tabular}{@{}lcccc|lcccc@{}}
\toprule
$K$ & 5 & 10 & 15 & 20 & $\tau$ & 5e-6 & 1e-5 & 2e-5 & 4e-5 \\ \midrule
CRS $\downarrow$ & 15.53 & 15.07 & 14.84 & 13.30 & CRS $\downarrow$ & 15.55 & 15.53 & 14.26 & 14.48 \\
Nudity $\downarrow$ & 0.02 & 0.02 & 0.01 & 0.01 & Nudity $\downarrow$ & 0.03 & 0.02 & 0.02 & 0.03 \\ \bottomrule
\end{tabular}
\end{table}
\begin{table}[htbp]
\centering
\caption{Results of LoRAShield adapted to SD3.}
\label{tab:sd3_results}
\setlength{\tabcolsep}{5pt}
\vspace{-10pt}
\tabcolsep=0.2cm
\renewcommand{\arraystretch}{0.8}
\aboverulesep=0ex
\belowrulesep=0.5ex
\begin{tabular}{@{}lccccc@{}}
\toprule
Dataset & CRS $\downarrow$ & IRS $\downarrow$ & BPS $\uparrow$ & FID $\downarrow$ & LPIPS $\downarrow$ \\ \midrule
Pixelart & 11.42 & 10.90 & 34.54 & 69.03 & 0.32 \\
3DM      & 14.33 & 11.20 & 32.76 & 55.47 & 0.34 \\
Trump    & 12.02 & 11.85 & 28.64 & 60.48 & 0.27 \\
Cat      & 10.43 & 10.63 & 31.89 & 75.08 & 0.35 \\ \bottomrule
\end{tabular}
\end{table}

\begin{proof}[Proof of Stability]
\label{app:proof}
Here we prove that tuning $W$ is stable while optimizing $B$ and $A$ is potentially unstable by analyzing their response to a gradient perturbation $\delta_g$. An update is stable if its response to $\delta_g$ is linear and constant, and unstable if amplified by a state-dependent factor.
For optimizing $W$, the parameter update is $\Delta W = -\eta \nabla_W L$. A perturbation $\delta_g$ to the gradient yields a change in the update $\Delta W' - \Delta W = -\eta \delta_g$. The magnitude of this change, $\lVert \Delta W' - \Delta W \rVert = \eta \lVert \delta_g \rVert$, is linearly proportional to the perturbation by a constant $\eta$, and is thus stable.
Conversely, for LoRA, the updates are $\Delta A = -\eta B^T \nabla_W L$ and $\Delta B = -\eta (\nabla_W L) A^T$. The same perturbation $\delta_g$ yields changes in the updates $\Delta A' - \Delta A = -\eta B^T \delta_g$ and $\Delta B' - \Delta B = -\eta \delta_g A^T$. The magnitudes are bounded by $\lVert \Delta A' - \Delta A \rVert \le \eta \lVert B^T \rVert \cdot \lVert \delta_g \rVert$ and $\lVert \Delta B' - \Delta B \rVert \le \eta \lVert A^T \rVert \cdot \lVert \delta_g \rVert$. Here, the response to the perturbation is amplified by the state-dependent norms of the trainable parameters, $\lVert A^T \rVert$ and $\lVert B^T \rVert$. This state-dependent amplification defines an unstable update process.
Thus, FFT is structurally stable, whereas LoRA is potentially unstable.
\end{proof}

Empirically, LoRAShield demonstrates the effectiveness of SVD. Theoretically, \textbf{the error introduced by Truncated SVD is controllable}, guaranteed by the Eckart-Young-Mirsky theorem. SVD is expressed as an outer product sum: $A = \sum_{i=1}^{r} \sigma_i u_i v_i^T$, where $\sigma_1 \ge \dots \ge \sigma_r > 0$ are the singular values, and $\{u_i\}$ and $\{v_i\}$ are the corresponding singular vectors. The optimal rank-$k$ approximation of $A$ is the truncated matrix $A_k$, which retains the first $k$ terms of this sum: $A_k = \sum_{i=1}^{k} \sigma_i u_i v_i^T$. The approximation error is: $A - A_k = \sum_{i=k+1}^{r} \sigma_i u_i v_i^T$. The magnitude of this error can be precisely quantified using the 2-norm of the error: $\|A - A_k\|_2 = \sigma_{k+1}$. These equations demonstrate that the approximation error is an explicit function of $\sigma_{k+1}$. 
\end{document}